# Ground state of $(Pb_{0.94}Sr_{0.06})(Zr_{0.530}Ti_{0.470})O_3$ in the morphotropic phase boundary (MPB) region: Evidence for monoclinic Cc space group

Ravindra Singh Solanki,[1] Akhilesh Kumar Singh,[1] S. K. Mishra,[2] Shane J. Kennedy,[3] Takashi Suzuki,[4] Yoshihiro Kuroiwa,[5] Chikako Moriyoshi,[5] and Dhananjai Pandey[1]

[1]School of Materials Science and Technology, Institute of Technology, Banaras Hindu University, Varanasi-221005, India

[2]Research and Technology Development Centre, Sharda University, Greater Noida-201306, India

[3]Australian Nuclear Science and Technology Organization (ANSTO), Australia

[4]Department of Quantum Matter, ADSM, Hiroshima University, Japan

[5]Department of Physical Science, Graduate School of Science, Hiroshima University, Japan

## Abstract

The antiferrodistortive (AFD) phase transition for a pseudotetragonal composition of $Pb(Zr_{0.530}Ti_{0.470})O_3$ (PZT) doped with 6% Sr has been investigated using sound velocity (4 to 320K), high resolution synchrotron X-ray powder diffraction (100 to 800K) and high resolution as well as high flux neutron powder diffraction measurements (4K) to settle the existing controversies about the true ground state of PZT in the morphotropic phase boundary (MPB) region. The multiplet character of the neutron diffraction profiles of $(3/2\ 1/2\ 1/2)_{pc}$ (pseudocubic or pc indices) and $(3/2\ 3/2\ 1/2)_{pc}$ superlattice peaks, appearing below the AFD transition temperature, rules out the rhombohedral R3c space group. The true ground state is confirmed to be monoclinic in the Cc space group in agreement with the predictions of the first principles calculations and earlier findings for pure PZT in the MPB region. 6% $Sr^{2+}$ substitution and the use of high wavelength (λ=2.44Å) neutrons have played key role in settling the existing controversies about the true ground state of PZT in the MPB region.

**Introduction:**

There is currently enormous interest in the study of composition induced morphotropic phase transition in a large variety of mixed perovskite solid solutions [1-6]. The morphotropic phase transition in ferroelectric systems is of immense technological significance, and forms the backbone of present day piezoelectric ceramic industry [7], as it is accompanied with an anomalous increase in the physical properties, like dielectric permittivity and piezoelectric coefficients, around the transition composition. However, there is no universal explanation for this enhancement of physical properties in different mixed solid solutions, although several models based on factors like phase coexistence [8], elastic instability [9-11], polarization rotation [6, 12 and 13] and polarization extension [14, 15] have been proposed. The relevance of morphotropic phase transition for high piezoelectric response was first discovered in the ferroelectric perovskite solid solution $Pb(Zr_xTi_{1-x})O_3$ (PZT) system [16-18] nearly six decades back. The phase diagram of PZT was shown to contain a morphotropic phase boundary (MPB) across which the structure changes from tetragonal symmetry for Ti-rich compositions to rhombohedral in the Ti-poor region [7]; the true symmetry of the rhombohedral phase is, however, now known to be monoclinic [19, 20].

While the MPB composition of PZT has been extensively exploited in a wide variety of commercial sensor and actuator devices, the physics behind the high piezoelectric response at the MPB is still not clear. The first step towards evolving a proper understanding of the physics of PZT ceramics is the knowledge of the ground state of PZT in the vicinity of the MPB. There is complete unanimity in the literature that the ground state of PZT in the MPB region is the result of a competition between ferroelectric and antiferrodistortive (AFD) instabilities associated with zone centre $\Gamma_4^-$ and zone boundary $R_4^+$ optical phonons of the paraelectric phase. Such a competition is known in other perovskite compounds and solid solutions also. For example, in $SrTiO_3$ the AFD instability freezes out at $T_c \sim 105K$ [21] and frustrates the subsequent development of ferroelectric (FE) order [22]. On the otherhand, in $BiFeO_3$ solid solutions it is now well documented that both the AFD and FE instabilities freeze out together as expected for a trigger type transition [23-25]. In contrast, in PZT, the FE instability freezes out first followed by the freezing out of the AFD instability [26, 27]. The AFD phase transition temperature of PZT, however, decreases on approaching the MPB [28] suggesting its weakening and eventual disappearance towards the Ti-rich side of the MPB.

While recent years have witnessed several interesting developments, both experimental and theoretical, related to the identification of the true ground state of PZT resulting from the freezing out of both ferroelectric and AFD instabilities near the MPB, there is still no unanimity about the correct space group of the ground state [29-35]. The first structural evidence for a new low temperature phase of PZT at the MPB composition (x=0.520) was reported by Noheda et al [29, 30] who discovered a tetragonal (P4mm space group) to monoclinic (Cm space group) phase transition below room temperature [30]. However, through a low temperature electron diffraction study of PZT, Ragini et al [26] showed that this monoclinic phase is not the ground state of PZT in the vicinity of the MPB. They found evidence for a yet another low temperature phase which is a superlattice phase stable below ~210K which cannot be accounted for in terms of the Cm space group reported by Noheda et al [29]. The existence of superlattice phase was further confirmed by neutron powder diffraction studies [27] which revealed that occurrence of an AFD phase transition involving antiphase tilting of oxygen octahedra at $T_c$~210K for x=0.520. The space group of the superlattice phase was initially proposed to be Pc [27] but was soon corrected as Cc [31]. It was also pointed out subsequently that this Cc phase coexists with a minority Cm phase [33]. The existence of the low temperature superlattice phase in the Cc space group was confirmed by other workers also in their TEM [36, 37] (for the compositions x=0.520 and 0.500, respectively) and powder neutron diffraction studies (for the composition x=0.520) as a function of temperature [38] and pressure [39].

Signatures of two low temperature transitions in the PZT with x=0.520, i.e. P4mm to Cm and Cm to Cc, are observed in the temperature dependence of physical properties and Raman spectra also. Ragini et al for PZT with x=0.515 and 0.520 [26] showed that the P4mm to Cm and the Cm to Cc phase transitions are accompanied with anomalies in the dielectric constant and elastic modulus, which was subsequently confirmed in dielectric and anelastic spectroscopic studies by other workers as well [10]. Raman scattering studies revealed new modes due to the P4mm to Cm phase transition [40] while resonance Raman scattering studies (Both for composition x=0.520) [41] confirmed the occurrence of the P4mm to Cm as well as Cm to Cc phase transitions, as evidenced by the appearance of new Raman lines below both the transition temperatures.

The experimental discovery of the monoclinic phase in the Cm space group was confirmed by first principles and phenomenological calculations also [6, 12 and 42] which also revealed a flattening of the free-energy profile in the MPB region as the principal mechanism of properties enhancement at the MPB through the rotation of the polarization vector [6]. In all these calculations, the AFD instability associated with the rotation/tilting of $BO_6$ octahedra was ignored eventhough it is known both experimentally [27] and theoretically [43] that there is an AFD instability in the vicinity of the MPB. A recent first principles work [34], which took into account both the AFD and the ferroelectric instabilities, suggests that the ground state of PZT near the MPB corresponds to the Cc space group and not Cm in agreement with the experimental observations [31, 33 and 38].

While the evidence in favour of the monoclinic Cm and Cc phases was mounting with Cc as the ground state of PZT in the MPB region, doubts were raised by some workers about the very existence of the monoclinic phases [32, 44 and 45]. Jin & Khachaturyan [44] and Wang [45] proposed that the so-called Cm phase is actually a consequence of miniaturised tetragonal or rhombohedral domains. However, such a theory cannot explain the appearance of anomalies in the temperature dependence of elastic modulus and dielectric constant at the two transitions temperatures. It also cannot account for the appearance of additional Raman lines below the two transition temperatures. Moreover, a recent convergent beam electron diffraction (CBED) study [46] has clearly revealed that the structure of the nanodomains in the MPB composition is indeed monoclinic. Frantti et al [32] argued that although there is an AFD transition in agreement with the previous workers [26, 27, 36], the ground state of PZT in the vicinity of the MPB is not monoclinic but rhombohedral in the R3c space group which coexists with the higher temperature Cm phase. While such a Cm to R3c phase transition cannot account for the additional Raman lines observed at low temperatures below the 'Cm to Cc' phase transition temperature [41], it has been claimed that the observed neutron powder diffraction profile of the superlattice phase of PZT in the vicinity of the MPB can be fitted to a model consisting of R3c and Cm space groups, of which the R3c phase was reported to be responsible for the superlattice reflections observed in the low temperature electron and neutron diffraction studies by various workers. However, it was shown subsequently that the consideration of coexistence of R3c and Cm phases leads to serious mismatch in the peak positions of the superlattice reflections in the Rietveld refinements using neutron powder diffraction data [33, 38]. But the controversy about the existence of the Cc phase has still not settled down as evidenced by recent publications [35, 47 and 48]. The main argument used against the Cc phase model is the non-observation of a superlattice reflection with $(1/2\ 1/2\ 1/2)_{pc}$ indices which is permitted in the Cc phase but is extinguished for the R3c phase [32]. Since the intensity of the $(1/2\ 1/2\ 1/2)_{pc}$ superlattice reflection is very low (~ 0.30% of the most intense peak), its observation above the background noise level remains a challenge till date. What then is the ground state of PZT? Is it Cc or R3c or none, since both the models are at the moment questionable, the former because of the non-observation of $(1/2\ 1/2\ 1/2)_{pc}$ superlattice peak [31] and the latter because of the misfit between observed and calculated profiles of some of the superlattice peaks [33, 38]. In this context, it would not be inappropriate to restate that the application of CBED technique of TEM for the determination of the symmetry of the lowest temperature phase of PZT in the MPB region has not revealed the existence of a phase in rhombohedral symmetry [46].

The present work was undertaken to settle the existing controversy about the space group (Cc vs R3c) of the true ground state of PZT ceramics by analyzing the profile shape of the most intense superlattice reflection $(3/2\ 1/2\ 1/2)_{pc}$ indices using high resolution neutron powder diffraction patterns collected at a high wavelength. This superlattice peak is a singlet for the R3c phase with (1 1 3) hexagonal indices, whereas it is a multiplet with four ((-3 1 2), (-2 2 1), (-1 1 2), (0 2 1)) monoclinic reflections nearly bunched together while another with indices (3 1 0) occurring away from this bunch. However, any meaningful analysis of the profile shape of even this strongest superlattice peak of PZT is fraught with several challenges such as its low intensity (~1.50% of the most intense peak) and its proximity to the most intense $(110)_{pc}$ peak leading to the overlap of the two profiles in all the experimental neutron powder diffraction patterns reported so far in the literature using low neutron wavelengths ($\lambda$=1.470, 1.540 and 1.667 Å by Frantii et al(2002) [32], Cox et al(2005) [38] and Ranjan et al(2005) [33], respectively). In order to analyse the profile shape of the $(3/2\ 1/2\ 1/2)_{pc}$ superlattice peak in an unambiguous manner, we have adopted two new strategies: (i) reduction of the average size of the A-site cation (i.e. $Pb^{2+}$, r=1.49Å) [49] in the $ABO_3$ structure by replacing it with 6% of smaller $Sr^{2+}$ (r=1.44Å) [49] ion which enhances the octahedral tilt angle and hence the intensity of all the superlattice peaks in general and the $(3/2\ 1/2\ 1/2)_{pc}$ superlattice peak in particular, and (ii) use of high wavelength neutrons such that the $(3/2\ 1/2\ 1/2)_{pc}$ peak is well separated from the neighbouring intense peak. Using these two strategies, we present unambiguous experimental evidence which proves that the correct space group of the ground state of PZT in the vicinity of the MPB is Cc and not R3c.

**Experimental and analysis:**

1. **Sample preparation:** Single phase powders of $(Pb_{0.94}Sr_{0.06})(Zr_xTi_{1-x})O_3$ (or simply PSZT) were obtained by solid state thermochemical reaction between

$(Pb_{0.94}Sr_{0.06})CO_3$ and $(Zr_xTi_{1-x})O_2$ for x=0.515, 0.520, 0.525, 0.530, 0.535, 0.545 and 0.550 at $800^0C$ for 6 hours. The $(Pb_{0.94}Sr_{0.06})CO_3$ precursor solid solution was synthesized by chemical co-precipitation technique to ensure homogeneous distribution of $Sr^{2+}$ at the $Pb^{2+}$ site. The $(Zr_xTi_{1-x})O_2$ solid solution was obtained from thermal decomposition of amorphous $(Zr_xTi_{1-x})(OH)_4$. $(Zr_xTi_{1-x})(OH)_4$ was prepared by co-precipitation technique to ensure unit-cell level homogenous distribution of $Zr^{4+}$ and $Ti^{4+}$ in the samples. Since our method of synthesis involves both solid state thermochemical reaction in a particulate mixture of two precursor solid solutions, which in turn were synthesized using chemical co-precipitation techniques, this approach has been termed as 'semi-wet' route in literature [50]. This route gives the narrowest width of the MPB region in PZT ceramics [28, 51]. The sintering of calcined powders has been carried out at $1100^0C$ for x≤0.535 and at $1050^0C$ for x≥0.545 in a sealed alumina crucible using $PbZrO_3$ as a spacer powder to prevent PbO loss. The sintered pellets were crushed to powder in an agate mortar and then annealed at $600^0C$ overnight to remove the strains introduced by crushing before subjecting the powder to diffraction studies.

**2. Characterization techniques:** High resolution Synchrotron X-ray powder diffraction (SXRD) measurements were carried out at room temperature and in the 100 to 800K range at BL02B2 beam line of SPring-8, Japan using a large Debye-Scherrer camera equipped with an imaging plate as a two-dimensional detector at a wavelength of 0.412Å (30 keV) [52]. High flux and high-resolution powder neutron diffraction data for $(Pb_{0.94}Sr_{0.06})(Zr_{0.530}Ti_{0.470})O_3$ (or simply PSZT530) were recorded using the powder neutron diffractometers Wombat and Echidna at ANSTO's OPAL facility [53, 54], respectively, at two wavelengths (1.66 and 2.44 Å). The Longitudinal elastic modulus ($C_L$) has been obtained by measuring the sound velocity (v) using the phase comparison-type pulse echo method [55-57]. The elastic modulus C was calculated using $C = \rho v^2$ with a room-temperature mass density ($\rho$) of the sintered sample. This measurement enabled us to locate the phase transition temperature.

**3. Profile refinement details:** Rietveld refinements for neutron powder diffraction data were carried out using the FULLPROF software package [58]. For the analysis of synchrotron X-ray powder diffraction (SXRD) data, we have used Le-Bail technique given in the FULLPROF package for determining the space group symmetry. We use linear interpolation for background fitting. A pseudo-Voigt peak shape function including anisotropic strain parameters has been chosen to generate profile shape for the peaks. The Cc space group has one Wyckoff site symmetry 4(*a*) with general coordinates. The unit cell consists of four formula units of PSZT530, and the asymmetric unit of the structure consists of five atoms: one Pb/Sr at (0.00, 0.75, 0.00), one Zr/Ti at (0.25+$\delta x_{Ti}$, 0.25+$\delta y_{Ti}$, 0.25+$\delta z_{Ti}$) and three oxygen atoms, O1 at (0.00+$\delta x_{O1}$, 0.25+$\delta y_{O1}$, 0.00+$\delta z_{O1}$), O2 at (0.25+$\delta x_{O2}$, 0.50+$\delta y_{O2}$, 0.00+$\delta z_{O2}$) and O3 at (0.25+$\delta x_{O3}$, 0.00+$\delta y_{O3}$, 0.50+$\delta z_{O3}$). The various $\delta$'s represent the refinable parameters. In keeping with the structural model used by Hatch et al [31], Pb was fixed at (0.00, 0.75, 0.00). There are four atoms in the asymmetric unit of the monoclinic phase with Cm space group. Pb/Sr was fixed at (0.00, 0.00, 0.00), Zr/Ti at (0.50+$\delta x_{Ti/Zr}$, 0.00, 0.50+$\delta z_{Ti/Zr}$), O1 at (0.50+$\delta x_{O1}$, 0.00, 0.00+$\delta z_{O1}$) and O2 at (0.25+$\delta x_{O2}$, 0.25+$\delta y_{O2}$, 0.50+$\delta z_{O2}$).

**Results and discussion:**

High resolution Synchrotron X-ray diffraction (SXRD) measurements were carried out on PSZT compositions with x=0.515, 0.520, 0.525, 0.530, 0.535, 0.545 and 0.550 to locate the morphotropic phase transition region. The evolution of the $(100)_{pc}$ and $(111)_{pc}$ perovskite reflections with composition is shown in Fig. 1. It is evident from these profiles that the $(100)_{pc}$ reflection is a doublet while $(111)_{pc}$ is a singlet for x=0.515. This is the characteristics of a tetragonal structure. At the other extreme of the composition, i.e. for x=0.550, the $(100)_{pc}$ reflection is apparently a singlet while $(111)_{pc}$ reflection is now a doublet. This is the characteristics of the so-called rhombohedral phase [51]. On increasing the Zr content from x=0.515, the width of the $(111)_{pc}$ peak starts increasing until for x=0.530 this peak splits and becomes an apparent doublet while $(100)_{pc}$ is still a doublet. This is the characteristics of the pseudotetragonal monoclinic phase in the Cm space group [19, 20]. With further increase in Zr content, the $(100)_{pc}$ peak splitting starts disappearing and it becomes a nearly 'singlet' but with a FWHM that is nearly 1.5 times that of the individual $(111)_{pc}$ peaks. This is the characteristics of the pseudorhombohedral monoclinic phase in the Cm space group [19, 20]. Thus, there is a morphotropic phase transition from the tetragonal structure to the 'pseudorhombohedral' monoclinic structure through a 'pseudotetragonal' monoclinic phase in the composition range 0.520≲x≲0.545. All the physical properties of PSZT are known to peak at x=0.530 [59] corresponding to the 'pseudotetragonal' monoclinic phase. This phase is the intermediate 'bridging phase' in the notation of Noheda et al [30] between pure tetragonal and pseudorhombohedral monoclinic phases for x≤0.515 and x≥0.550, respectively. As in the case of pure PZT, we have found evidence for the coexistence of phases of neighbouring single phase compositions in the morphotropic phase transition region. We shall discuss the phase coexistence aspect in relation to the MPB composition of PSZT (i.e. x=0.530) in the next section. The fact that SXRD profiles change significantly with as small a change in composition as Δx=0.005 (compare e.g. the profiles for x=0.525 and 0.530 in Fig. 1) proves that PSZT samples used in the present investigation, prepared by semi-wet route described in relation to the synthesis of pure PZT [50], possess excellent chemical homogeniety. The chemical homogeneity of PSZT samples was further confirmed by Williamson-Hall (W-H) analysis of the SXRD profiles in the high temperature cubic phase. Fig. 2 depicts the W-H plot for PSZT with x=0.530 in the cubic phase at T=800K. The value of Δd/d, which is a measure of compositional fluctuation in the sample, is found to be ~$3\times10^{-4}$ which is comparable to pure PZT samples (see ref. 29 for pure PZT with Δd/d ~$3\times10^{-4}$ in the cubic phase). We can thus conclude that $Sr^{2+}$ substitution does not lead to any large scale segregation in our PSZT samples prepared by the semi-wet route in marked contrast to the PSZT samples prepared by solid solution route [60]. It is interesting to note that the MPB of pure PZT corresponds to a peak in $d_{33}$ and other physical properties that occurs at x=0.520 [7, 51] while it shifts to x=0.530 as a result of 6% $Sr^{2+}$ substitution [59]. We have accordingly concentrated on 6% $Sr^{2+}$ substituted PZT with x=0.530 i.e. $(Pb_{0.94}Sr_{0.06})(Zr_{0.530}Ti_{0.470})O_3$ in the present investigation for studying the AFD phase transition corresponding to the MPB composition.

Fig. 3 depicts the evolution of SXRD profiles of the $(111)_{pc}$, $(200)_{pc}$ and $(220)_{pc}$ reflections on cooling from the paraelectric cubic phase down to T=100K. It is evident from this figure that at and above 600K, all the profiles are singlet. On cooling below 600K, the (200) and (220) cubic peaks split into two peaks while (111) remains a singlet. As discussed earlier, this is the characteristics of a tetragonal phase. On cooling below 425K, the singlet $(111)_{pc}$ peak also splits while the weak reflection of the $(220)_{pc}$ profile shifts from higher 2θ angle at 425K to lower 2θ angle at 400K. In addition, there is another hump on the higher 2θ angle of the (220)$_{pc}$ profile. The (200)$_{pc}$ profile, on the other hand, continues to exhibit doublet character. All these features clearly indicate a structural phase transition to the monoclinic phase in the Cm space group, as was first discovered by Noheda et al [29] in PZT with x=0.520. On cooling the sample below 300K, there is no noticeable change in the shape of the SXRD profiles as shown in Fig. 3 which is similar to what was observed by Noheda et al [30] and Ragini et al [26] below the tetragonal (P4mm) to monoclinic (Cm) phase transition temperature of ~250K in pure PZT with x=0.520.

In order to confirm the change in space group symmetries as a function of temperatures, we depict in Fig. 4 the Le-Bail fits of selected perovskite reflections (111)$_{pc}$, (200)$_{pc}$ and (220)$_{pc}$ using SXRD data for the Pm$\bar{3}$m space group at 800K, P4mm space group at 550K, Cm & (Cm+Cm) space groups at 300K and (Cm+Cm) space groups at 100K. From Fig. 4(c), which depicts the fits using single Cm phase, it is evident that there is mismatch in the observed (red dots) and calculated (black line) pattern for the (002)$_{pc}$ & (200)$_{pc}$ reflections. It was found necessary to consider another monoclinic phase (a minority phase) in Cm space group to account for this mismatch. This phase may be a coexisting phase corresponding to a nearby composition due to small compositional fluctuations. On the same logic, the refinements at 100K were carried out using two Cm phases. The consideration of the minority Cm phase leads to significant improvement in the overall fit as evidenced by a drastic reduction in $\chi^2$ also at 300K. The presence of a minority Cm phase was further confirmed by Rietveld refinements at 300 and 100 K. The phase fraction of this minority phase comes out to be 24% at 300K and does not show any significant systematic variation with decreasing temperature. However, we would like to mention that in view of the large anisotropic peak broadening of the minority Cm phase as compared to that of the majority Cm phase, the phase fractions can not be obtained reliably from the integrated intensities. The real phase fraction of the Cm phase may in fact be smaller than the values obtained by the Rietveld refinements.

As pointed out in the introduction section, the phase transition from the monoclinic phase in the Cm space group to the lower temperature superlattice phase of PZT is an antiferrodistortive phase transition involving anti-phase rotation of oxygen octahedra. Such an antiferrodistortive phase transition affects only the oxygen positions and that too rather slightly. As is well known, XRD technique is not sensitive to small changes in oxygen positions on account of its low atomic number. In order to capture the signatures of this antiferrodistortive phase transition, electron and neutron diffraction data are required as was first pointed out by Ragini et al [26] and Ranjan et al [27] in the context of PZT. Fig. 5 compares the medium resolution but high flux neutron powder diffraction pattern (a) of the PSZT530 at 100K with the SXRD pattern (b), which clearly shows the presence of several superlattice reflections (marked with arrows) which are not discernible in the corresponding SXRD pattern shown in Fig. 5(b). All the main perovskite peaks and the superlattice reflections could be indexed using a pseudocubic perovskite cell. With reference to such a cell, the perovskite peaks acquire integral pseudocubic Miller indices while superlattice reflections assume fractional Miller indices involving three odd (o) integers. Fig. 6 depicts the evolution of the neutron powder diffraction profiles of $(1/2\ 1/2\ 1/2)_{pc}$, $(3/2\ 1/2\ 1/2)_{pc}$, $(3/2\ 3/2\ 1/2)_{pc}$ and $(5/2\ 1/2\ 1/2)_{pc}$ superlattice reflections as a function of temperature. It is evident from this figure that the intensity of the superlattice reflections decreases drastically on increasing the temperature but remains non-zero even at room temperature. These observations clearly show that the monoclinic Cm phase of PSZT530 undergoes an antiferrodistortive phase transition into a superlattice phase which is not revealed by the SXRD data shown in Figs. 3 and 5(b).

Additional support for the AFD transition was obtained by studying the temperature dependence of the longitudinal ultrasonic velocity. Fig. 7 depicts the longitudinal elastic modulus, obtained from the measured ultrasonic velocity and density of the sample, as a function of temperature. For a normal solid which contracts on cooling, the elastic modulus should increase with decreasing temperature as it is inversely proportional to the interatomic distance:

$$C = \frac{1}{r_0} \left( \frac{\partial^2 U}{\partial r^2} \right)_{r=r_0}$$

where U is the interatomic potential and $r_0$ is the equilibrium interatomic distance. However, the longitudinal elastic modulus of PSZT530 shows an anomalous behaviour in the sense that it decreases with decreasing temperature upto 260K. Such an anomalous behaviour occurs as a result of lattice instability due to soft modes as a precursor to an impending structural phase transition. Interestingly, below 260K the temperature dependence of the longitudinal modulus becomes normal suggesting a transition temperature $T_c \sim 260K$ for the AFD phase transition. The hysteresis in the transition temperature, shown more clearly in the inset, measured during heating and cooling cycles, indicates the first order character of this transition. Since first order phase transitions are accompanied with coexistence of the high and low temperature phases across the phase transition temperature, we believe that the intensity of the superlattice reflections does not become zero above AFD transition temperature $T_c \sim 260K$.

In order to address the existing controversy about the correct space group of the superlattice phase, we collected high resolution neutron powder diffraction data using ECHIDNA diffractometer at two different neutron wavelengths: 1.66Å and 2.44Å. Fig. 8 compares the neutron powder diffraction patterns of PSZT530 at 300 and 4K for the two neutron wavelengths. The previous workers recorded neutron powder diffraction data at smaller wavelengths ($\lambda$=1.470, 1.667 and 1.540 Å by Frantii et al(2002) [32], Ranjan et al(2005) [33] and Cox et al(2005) [38] respectively) only. However, for smaller wavelengths, the diffraction profile of the $(3/2\ 1/2\ 1/2)_{pc}$ superlattice reflection, which is the strongest in intensity amongst all the superlattice reflections, is partly overlapping with the neighbouring intense $(110)_{pc}$ perovskite peak as shown in the inset of Fig. 8(b). As a result, its profile shape cannot be analysed reliably using smaller wavelength data. Use of higher neutron wavelength enables reasonable separation of the profile of the $(3/2\ 1/2\ 1/2)_{pc}$ superlattice peak from the neighbouring intense perovskite peak, as can be seen from the inset of Fig. 8(d). Fig. 9(a) gives a magnified view of the $(3/2\ 1/2\ 1/2)_{pc}$ superlattice reflection. It is evident from the profile shape of this superlattice peak that it cannot be a singlet. It can easily be deconvoluted into two peaks. The $(3/2\ 1/2\ 1/2)_{pc}$ superlattice reflection for the R3c space group is a singlet consisting of (1 1 3) (hexagonal indices) reflection only. On the other hand, for the Cc space group this superlattice peak is a multiplet consisting of (-3 1 2), (-2 2 1), (-1 1 2), (0 2 1) and (3 1 0) (monoclinic indices) reflections. Of these, the (-3 1 2), (-2 2 1), (-1 1 2), and (0 2 1) reflections are nearly bunched together while (3 1 0) occurs at a lower 2θ angle as shown in Fig. 9(b) which depicts the Rietveld fit around this reflection. It is because of this, the overall profile shape of the $(3/2\ 1/2\ 1/2)_{pc}$ reflection appears as a doublet in Fig.9. The doublet nature of the $(3/2\ 1/2\ 1/2)_{pc}$ superlattice peak clearly rules out the possibility of the R3c space group for the superlattice phase of PSZT530. A perusal of the profile of the $(3/2\ 3/2\ 1/2)_{pc}$ superlattice peak, which has smaller intensity than the strongest $(3/2\ 1/2\ 1/2)_{pc}$ superlattice peak, in Fig. 6 clearly reveals that even this peak is not a singlet. For rhombohedral R3c space group, this peak should also have been a singlet. What about the $(1/2\ 1/2\ 1/2)_{pc}$ peak which has non-zero intensity for the Cc space group and is extinguished for the R3c space group? In Fig. 6, we clearly see a peak at $2\theta \sim 30.87^0$ (corresponding to $d_{hkl}=4.59$Å) which nearly corresponds to the $(1/2\ 1/2\ 1/2)_{pc}$ superlattice peak. However, the intensity of this superlattice peak does not decrease much, unlike the intensity of all other superlattice peaks like $(3/2\ 1/2\ 1/2)_{pc}$, $(3/2\ 3/2\ 1/2)_{pc}$ and $(5/2\ 1/2\ 1/2)_{pc}$, on increasing the temperature from 4K to 300K. This suggests that in addition to the $(1/2\ 1/2\ 1/2)_{pc}$ peak of the Cc space group, there is some spurious contribution from extraneous sources around the same $2\theta$ angle. Non-PZT samples also sometimes showed a small peak at this position using the same cryostat for changing the sample temperature. This led us to conclude that some spurious contribution from the cryostat occurs at the same $2\theta$ angle as expected for the $(1/2\ 1/2\ 1/2)_{pc}$ superlattice peak. In view of this, we did not analyse the intensity near $(1/2\ 1/2\ 1/2)_{pc}$ position further. But eventhough this reflection is masked by spurious contributions, the non-singlet nature of the $(3/2\ 1/2\ 1/2)_{pc}$ and $(3/2\ 3/2\ 1/2)_{pc}$ (Fig. 6) superlattice peaks is sufficient to rule out the R3c space group atleast for PSZT530 samples. This leaves behind the Cc space group as the other plausible space group for the low temperature superlattice phase of PSZT530.

In order to confirm the correctness of the Cc space group for the low temperature superlattice phase of PSZT530, we carried out Rietveld analysis using both the SXRD and neutron powder diffraction data, the latter data for λ=2.44Å. Since the neutron powder diffraction pattern recorded at ECHIDNA using high neutron wavelength contains very few peaks, we decided to determine the Zr/Ti positional corrdinates using SXRD data by Rietveld refinement considering two coexisting monoclinic phases Since the neutron scattering lengths of Ti and Zr are of opposite sign leading to a very small value of the Ti/Zr site scattering length, the location of the atoms may be made more reliably using SXRD. These coordinates were used as the input parameters for the Rietveld refinements using neutron data for the Cc space group. In the refinements using neutron powder diffraction data, we found that the consideration of only one monoclinic phase with Cc space group gives fairly good fit except for the $(h00)_{pc}$ type reflections. This is illustrated in the left hand column of Fig. 10. As mentioned earlier, we had a similar situation with the Rietveld fits using SXRD data. The consideration of the minority coexisting Cm phase led to an overall improvement in the fits with value of $\chi^2$ reducing from 3.60 to 2.96. The phase fraction of the Cc and Cm phases comes out to be 70% and 30%, respectively, which are close to the values obtained from XRD data keeping in mind the difference in the anisotropic peak broadening parameters for the two data. It is evident from the fits of the three superlattice peaks that none of these is a singlet which comprehensively rules out the R3c space group. During the refinements, we first used constrained model proposed by Ranjan et al [27] that involves rigid body rotation of oxygen octahedra leading to only five refinable oxygen positional coordinates but it was found that the constrained model did not give good fit and invariably led to higher $\chi^2$ values. So, unconstrained model considering refinement of the entire space group permitted oxygen positions and all three Zr/Ti positional coordinates was used. It is found that unconstrained model gives overall good fit and reasonable agreement factors, except that the thermal parameter for Zr/Ti was converging to very small but negative value which is physically unrealistic. Accordingly this parameter was arbitrarily kept fixed at a small positive value close to zero during refinements. Table 1 gives the values of refined parameters and agreement factors for PSZT530 at 4K for coexisting Cc and Cm phases. Fig. 11 depicts the full pattern Rietveld fits using neutron powder diffraction data at 4K for the Cc+Cm phase model. The fits are evidently quite good. We thus conclude that the ground state of PSZT530 in the MPB region does not correspond to the R3c space group but to the monoclinic Cc space group.

**Concluding remarks:**

The present dispute about the ground state of PZT has two facets: (i) what is the space group of the ground state of PZT below the AFD transition temperature for pseudotetragonal compositions in the MPB region and (ii) what is the space group below the AFD transition temperature for the pseudorhombohedral compositions outside the MPB region. In the present work, we have attempted to resolve the first controversy by doping PZT with 6% $Sr^{2+}$ which enhanced the intensity of the main superlattice peak (3/2 1/2 1/2)$_{pc}$, and by using high wavelength neutrons which enabled separation of the (3/2 1/2 1/2)$_{pc}$ peak from the neighbouring, intense (111)$_{pc}$ peak. The main finding of the present work is that as a result of the AFD phase transition for the MPB composition of such a lightly doped PZT composition, the structure of the monoclinic phase changes from Cm space group to the Cc space group just below room temperature. The R3c space group postulated by some workers [32, 35, 44 and 45] as the true ground state of PZT can be rejected for the MPB composition of PSZT, as it cannot account for the multiplet character of the superlattice peaks like $(3/2\ 1/2\ 1/2)_{pc}$ and $(3/2\ 3/2\ 1/2)_{pc}$. Our results are in agreement with a host of experimental findings on MPB compositions of PZT [31, 37-39] as also the predictions of the first principle calculations [34] for the ground state of PZT.

As regards the second controversy, the AFD phase transition has been known in pure PZT ever since its phase diagram above room temperature was established [7]. As per the old phase diagram, the higher temperature rhombohedral phase of PZT in the R3m space group transforms to another rhombohedral phase in the R3c space group as a result of an AFD phase transition above room temperature in the composition range $0.620 \lesssim x \lesssim 0.940$. Recent studies based on high resolution SXRD as well as rotating anode data on 'rhombohedral' compositions of R3m space group of PZT have revealed that even these compositions are monoclinic in Cm space group at a medium range length scale [11, 19]. Corker et al [61] had in an early study shown that in the Rietveld refinements of PZT using neutron powder diffraction data for $0.60 \leq x \leq 0.88$, Pb is displaced locally in the $<110>_{pc}$ direction, which essentially conforms to a local monoclinic distortion of the structure [62]. This suggested that the rhombohedral structure with R3c space group of PZT could also be medium or short range ordered monoclinic structure [28]. Based on previous studies by Ragini et al [19], Singh et al [20], Corker et al [61], Glazer et al [62] and Pandey et al [28] proposed that the ground state space group of PZT in the composition range $0.620 \lesssim x \lesssim 0.940$ should also be Cc. However, in a recent single crystal study on PZT compositions with x=0.540 and 0.675, it has been argued that the space group of PZT below AFD transition temperature for such compositions is R3c and not Cc, since they did not observe any intensity at the $(1/2\ 1/2\ 1/2)_{pc}$ position [35]. In the present work, we have addressed the issue of the space group of the ground state of pseudotetragonal compositions of PSZT in the MPB region. Our results clearly show that atleast for the MPB composition of the PSZT system, the ground state corresponds to a monoclinic phase in the Cc space group and not the R3c space group. Our future efforts will now be directed towards the identification of the true ground state of pseudorhombohedral compositions of 6% $Sr^{2+}$ substituted PZT away from the MPB.

**Acknowledgements:**

R. S. Solanki acknowledges financial support from Council of Scientific and Industrial Research (CSIR), India in the form of a Junior Research Fellowship. D. Pandey and Y. Kuroiwa acknowledge financial support from Department of Science and Technology (DST), Govt. of India and Japan Society for the Promotion of Science (JSPS) of Japan under the Indo-Japan Science Collaboration Program. The synchrotron radiation experiments were performed at the BL02B2 of Spring-8 with the approval of Japan Synchrotron Radiation Research Institute (Proposal Nos. 2011A1324 and 2011A0084). We acknowledge the assistance of A. Studer and J. Hester from ANSTO, Australia for neutron data collection and I. Ishii and H. Muneshige of the Department of Quantum Matter, Hiroshima University, Japan for ultrasonic measurements.

**Figure Captions:**

**Figure 1.** Synchrotron powder XRD profiles of the $(100)_{pc}$ and $(111)_{pc}$ peaks of PSZT for (a) x=0.515, (b) x=0.520, (c) x=0.525, (d) x=0.530, (e) x=0.535, (f) x=0.545, (g) x=0.550.

**Figure 2.** Williamson-Hall plot for the cubic phase of PSZT530 at 800K.

**Figure 3.** (Color online) The evolution of synchrotron powder XRD profiles of the $(111)_{pc}$, $(200)_{pc}$ and $(220)_{pc}$ reflections of PSZT530 with temperature.

**Figure 4.** (Color online) Observed (dots), calculated (continuous line), and difference (bottom line) profiles of selected $(111)_{pc}$, $(200)_{pc}$ and $(220)_{pc}$ reflections at different temperatures obtained after Lebail fitting. The vertical tick marks above the difference profiles give the positions of the Bragg reflections.

**Figure 5**. Powder diffraction patterns of PSZT530 at 100K: (a) medium resolution high flux neutron data recorded at λ=2.440 Å and (b) high resolution synchrotron data at λ=0.412 Å. The superlattice reflections with fractional indices are clearly seen in (a).

**Figure 6.** (Color online) The evolution of powder neutron profiles of the $(1/2\ 1/2\ 1/2)_{pc}$, $(3/2\ 1/2\ 1/2)_{pc}$, $(3/2\ 3/2\ 1/2)_{pc}$ and $(5/2\ 1/2\ 1/2)_{pc}$ superlattice reflections of PSZT530 as a function of temperature. This is medium resolution but high flux data recorded at λ=2.44Å using WOMBAT diffractometer.

**Figure 7.** Temperature variation of longitudinal elastic modulus ($C_L$) of PSZT530. Inset shows a magnified view of the AFD phase transition region showing hysteresis.

**Figure 8.** (Color online) High resolution neutron powder diffraction patterns of PSZT530 at 300 and 4K collected using neutron wavelengths 1.66Å (a, b) and 2.44Å (c, d), respectively, on the ECHIDNA diffractometer. Insets (a, c) show the near absence of superlattice (3/2 1/2 1/2)$_{pc}$ at 300K and insets (b, d) show the zoomed view of superlattice (3/2 1/2 1/2)$_{pc}$ at 4K. Note the sepaeration of the peak from the neighbouring perovskite peak in the inset of (d) for λ=2.440Å.

**Figure 9.** (Color online) (a) Deconvolution of the profile of (3/2 1/2 1/2)$_{pc}$ superlattice reflection obtained using high wavelength, high resolution neutron diffraction data at 4K. (b) Rietveld fit of (3/2 1/2 1/2)$_{pc}$ superlattice reflection using Cc space group.

**Figure10.** (Color online) Observed(dots),calculated(continuous line) and difference (bottom curve) plots of the various superlattice peaks and three perovskite peaks (111)$_{pc}$, (200)$_{pc}$ and (220)$_{pc}$ of PSZT530 at 4K, as obtained by Rietveld refinement. The left coloumn gives the fits for the Cc space group while the right column gives fits for coexisting Cc and Cm space groups. The neutron diffraction data was taken on ECHIDNA diffractometer.

**Figure 11.** (Color online) The overall fit between observed and calculated patterns obtained by Rietveld refinement using Cc+Cm space group at 4K. The neutron diffraction data was taken on ECHIDNA diffractometer.

**Table captions:**

**Table 1.** Refined structural parameters and agreement factors for the coexisting Cc and Cm space groups of PSZT530 at 4K.

**Table1**

| | Space group: Cc | | | | Space group: Cm | | | |
|---|---|---|---|---|---|---|---|---|
| Atoms | x | y | z | B (Å$^2$) | x | y | z | B (Å$^2$) |
| Pb/Sr | 0.00 | 0.75 | 0.00 | $\beta_{11}$=0.0091(9) | 0.00 | 0.00 | 0.00 | $\beta_{11}$=0.016(3) |
| | | | | $\beta_{22}$=0.008(3) | | | | $\beta_{22}$=0.033(3) |
| | | | | $\beta_{33}$=0.025(4) | | | | $\beta_{33}$=0.036(5) |
| | | | | $\beta_{13}$=0.010(1) | | | | $\beta_{13}$=0.002(2) |
| Ti/Zr | 0.236(1) | 0.269(5) | 0.271(4) | 0.001 | 0.500(3) | 0.00 | 0.418(3) | 0.003 |
| O1 | -0.0499(8) | 0.2663(2) | 0.027(2) | 1.6(4) | 0.567(1) | 0.00 | -0.062(2) | 0.589(1) |
| O2 | 0.2103(9) | 0.507(2) | 0.008(2) | 0.05(1) | 0.299(1) | 0.2633(8) | 0.401(1) | 0.685(2) |
| O3 | 0.1821(7) | -0.024(2) | 0.442(2) | 0.904(1) | | | | |

a=10.0343(8)Å, b=5.711(1)Å, c=5.725(2)Å    a=5.735(5)Å, b=5.729(4)Å, c=4.094(4)Å

$\beta$=125.22$^0$(6)                                                   $\beta$=90.50$^0$(3)

Weight fraction           0.70                                    0.30

$R_{wp}$ = 4.47, $R_P$=3.33, $R_{exp}$= 2.60, $\chi^2$=2.96

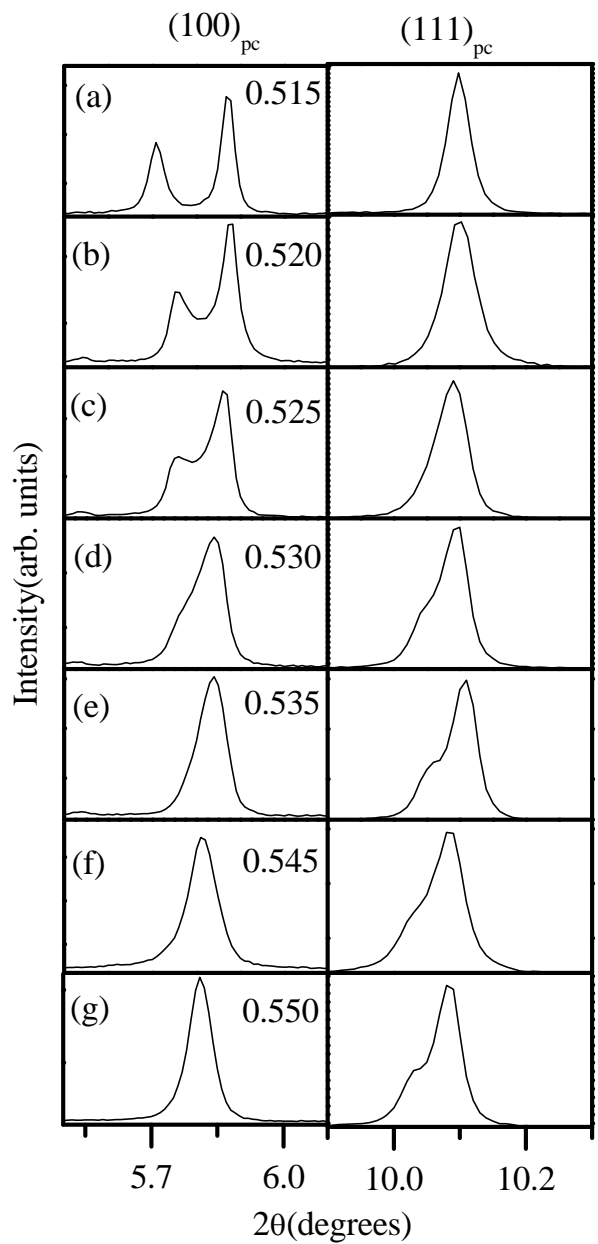
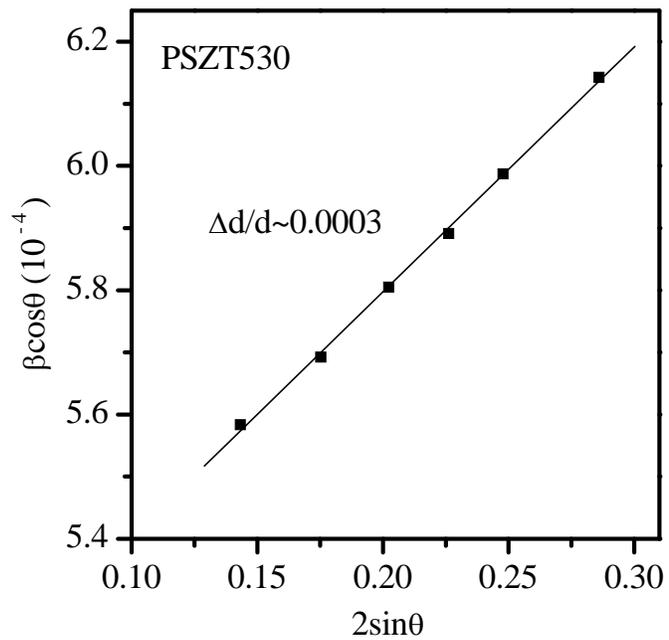

Fig.1

Fig. 2

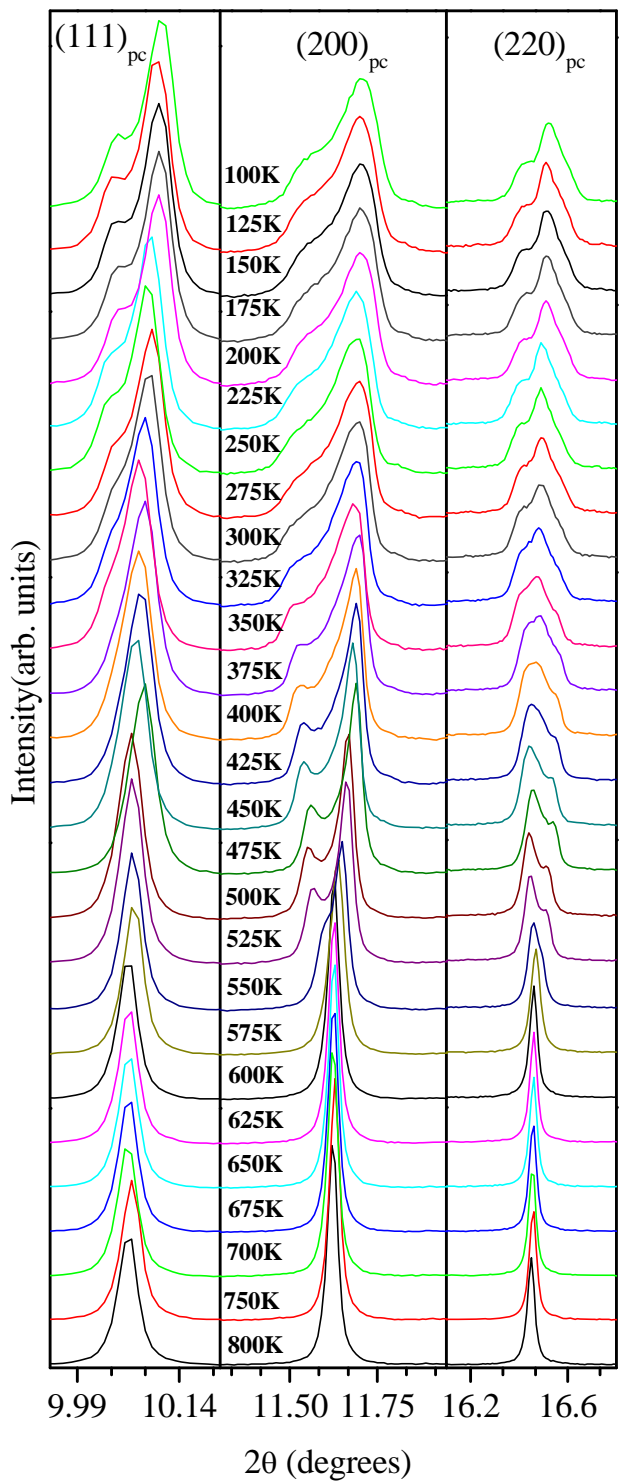

Fig. 3

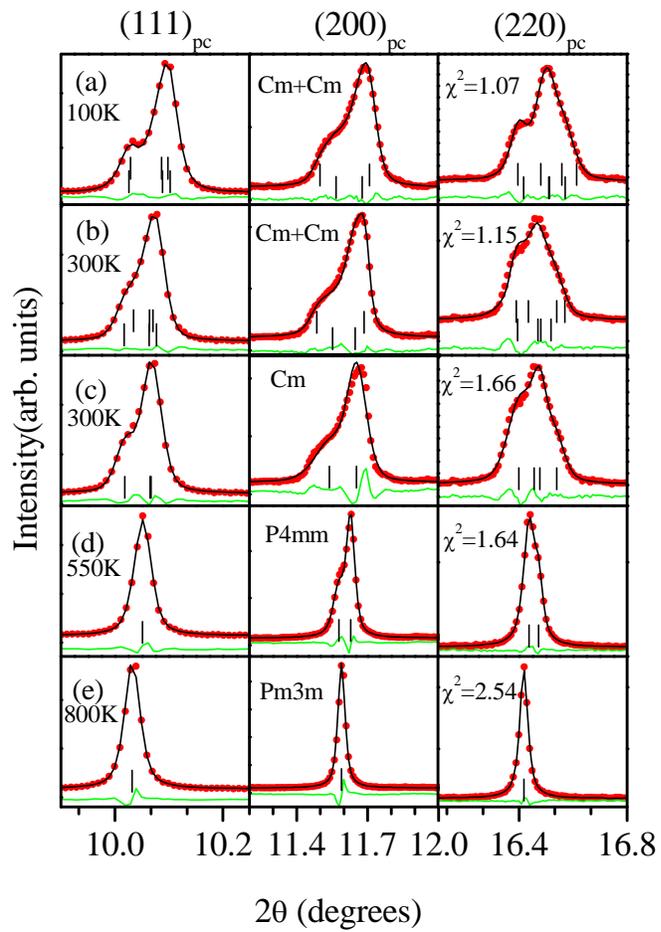

Fig.4

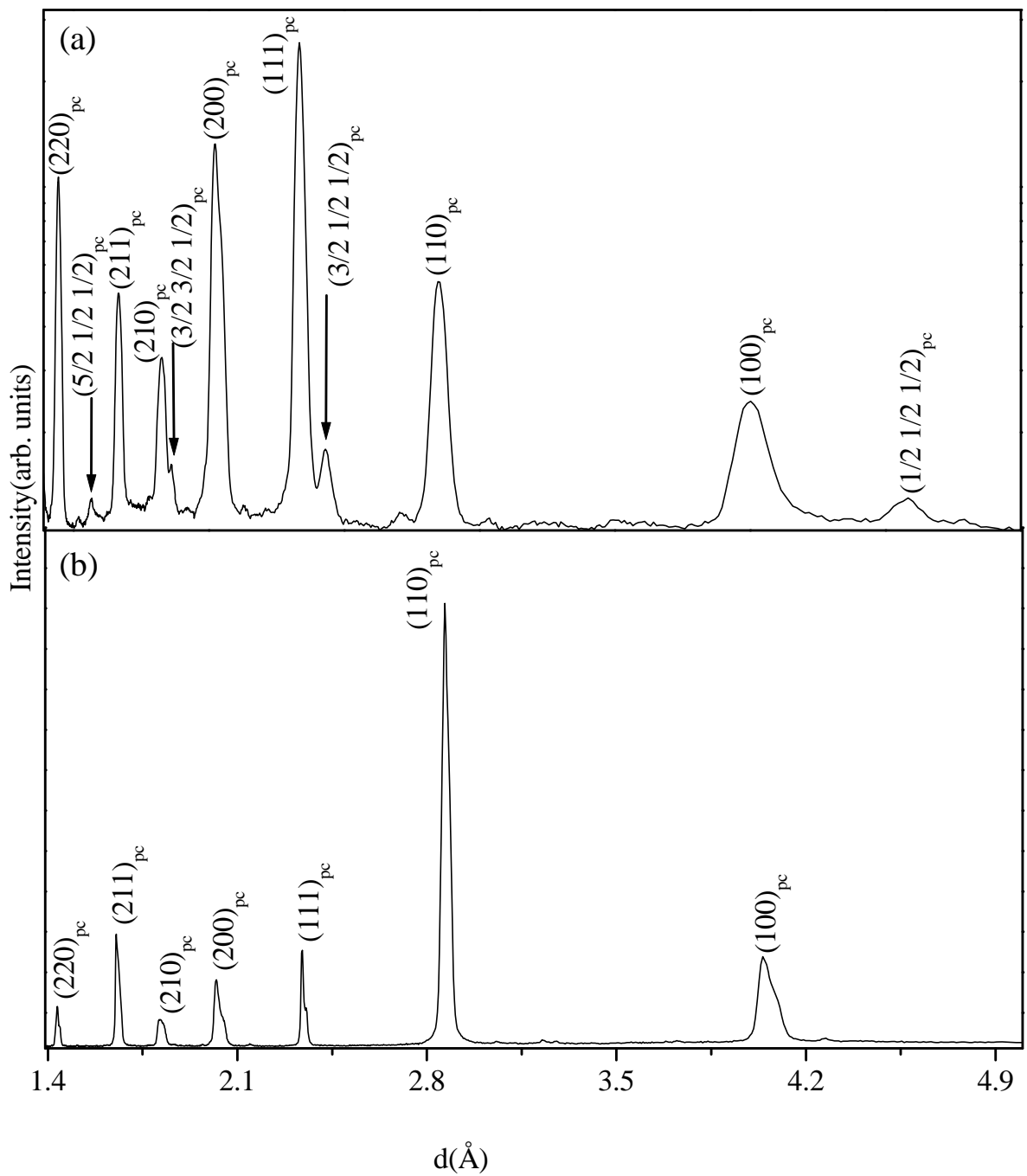

Fig.5

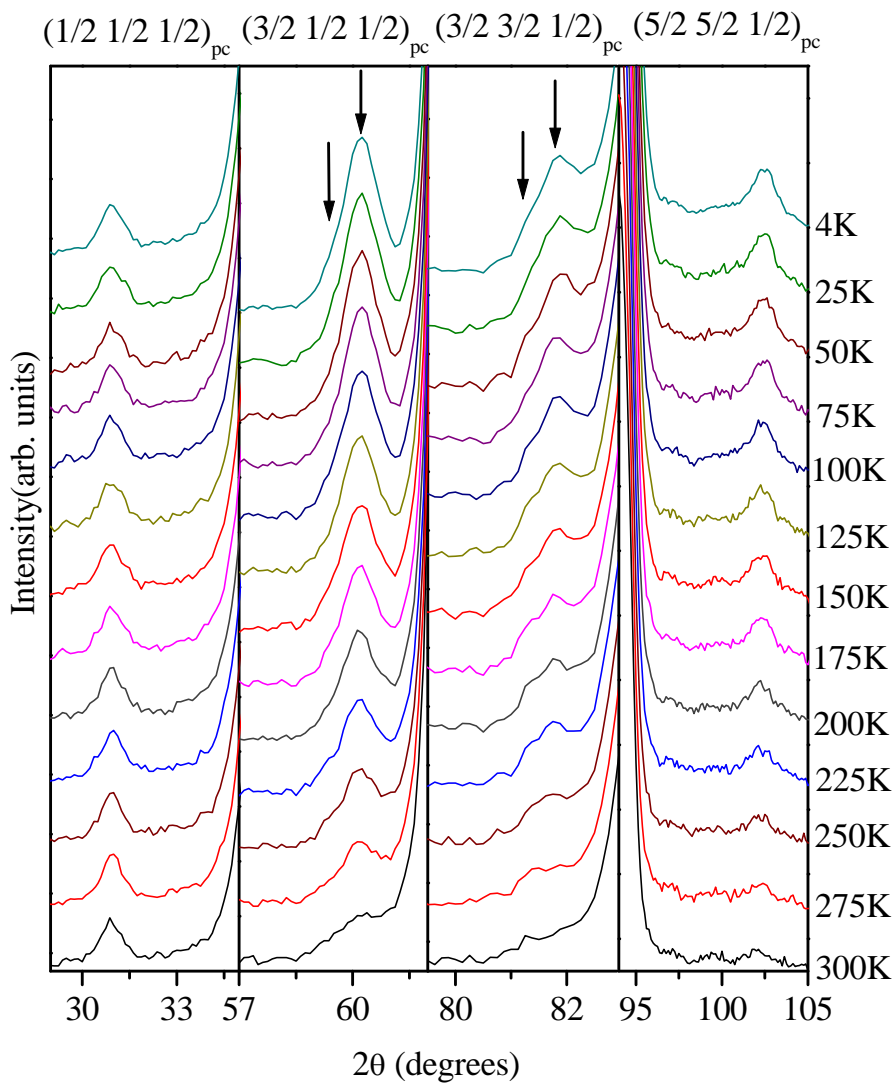

Fig.6

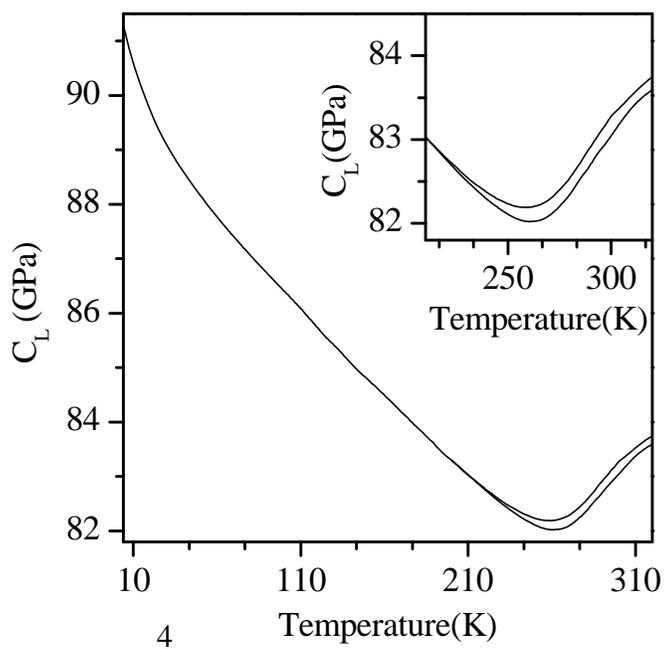

Fig.7

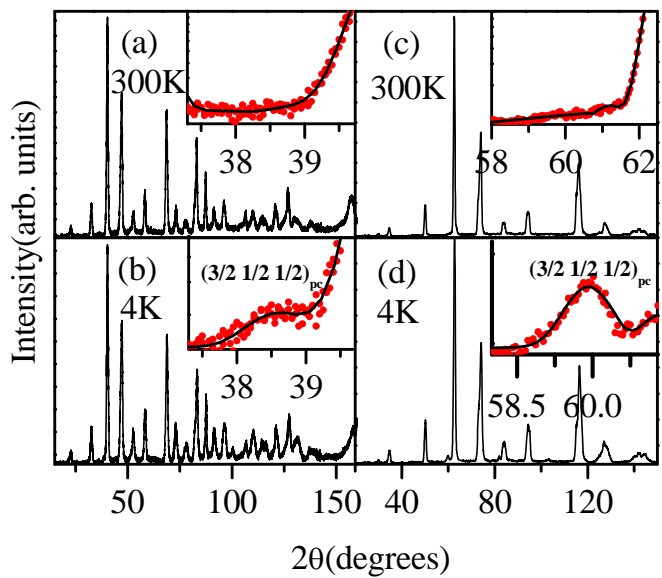

Fig.8

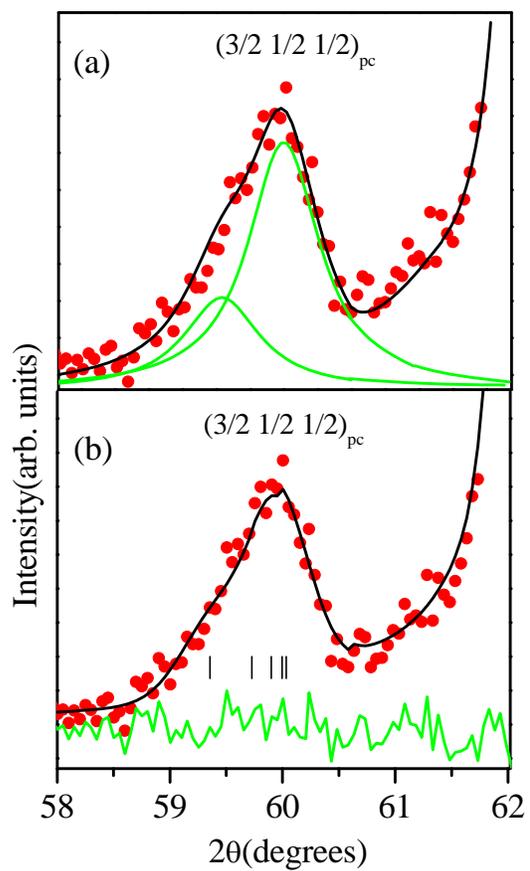

Fig.9

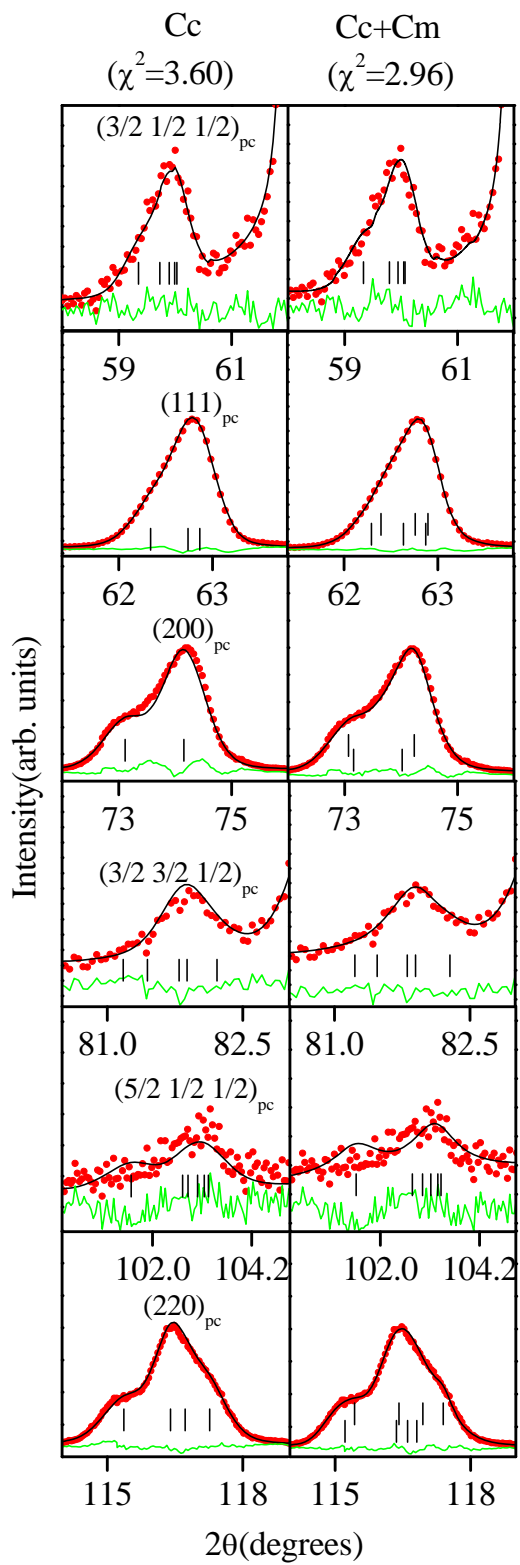

Fig.10

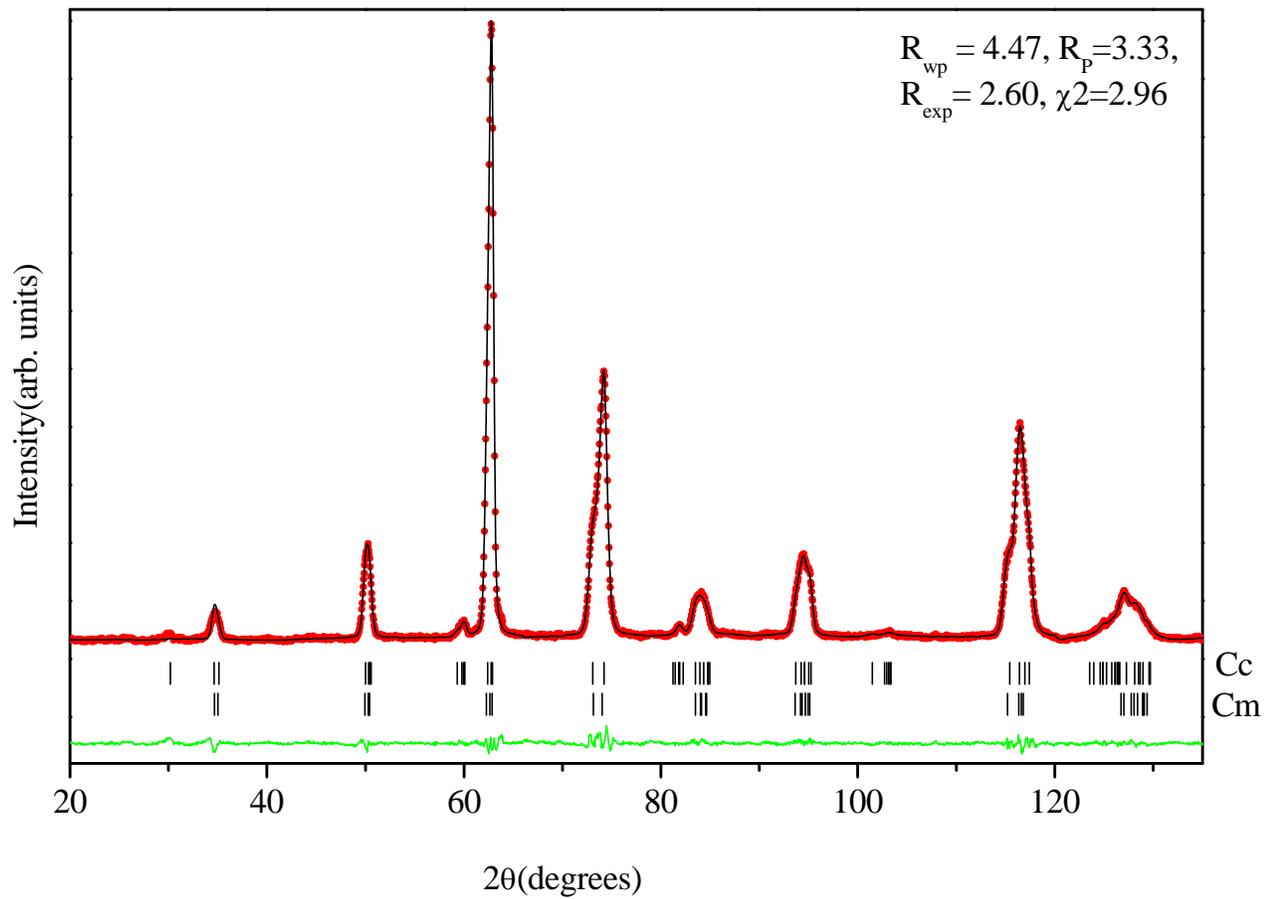

Fig.11